\documentclass[twocolumn,aps,prl,showpacs,superscriptaddress]{revtex4}
\usepackage[dvips]{graphicx}
\usepackage{bm}

\begin{document}
\title{Surface defects in vesicle aggregates and anyons}

\author{L.V. Elnikova}
\affiliation{
A.I.~Alikhanov Institute for Theoretical and Experimental Physics, \\
B. Cheremushkinskaya st., 25, 117218 Moscow, Russia}
\date{\today}

\begin{abstract}
We consider surface defects in connection to the closed vesicle
form evolution in mesomorphism of lyotropic aggregates, based on
the experimental data by Feigenson \cite{Feig} on confocal
fluorescent resonant microscopy for the lipid $DPPC/
DLPC$/cholesterol system. To estimate the influence of surface
topological defects onto aggregate form it has been used the
fractional quantum Hall effect (FQHE) description \cite{EPL1992},
\cite{Haldane1}.
\end{abstract}


\maketitle
 \textbf{1. Introduction.} Topological methods are suitable for the
description of surface and bulk defects \cite{MM}, \cite{Volovik},
and also for describing the shape evolution of lyotropic liquid
crystal aggregates. Some types of surface defects of micelles and
vesicles with closed surface have been predicted in frame of
fluctuation-dissipation (FD) theorem and Goldstone theorem
\cite{Milner}. This studying stimulated by observed phenomena of
surfaces of nanoscopic domains in lipid mixture
dipalmitoylphosphatidylcholine/dilauroylphosphatidylcholine/cholesterol
($DPPC/DLPC/$cholesterol) \cite{Feig} that have not topological
interpretation. In Feigenson experiments \cite{Feig} on lyotropic
mesomorphism it was observed phase inhomogeneity at asymmetric
shape of closed lipid vesicle, which also may be considered as
$n$-axis gyration effects induced by the mixture component
variation. Minimum surface domains diameter is on the order of
$50-60 \AA$ and 
an island consist about 40-60 lipids in 
1-phase regime (D - region at the phase diagram, Fig. 1, Fig. 2.)
between the different cholesterol concentration.

The widespread description for phase behavior of fluid membranes
is based on the Ginzburg-Landau (GL) free-energy functional. As is
well known, this functional with the same the order parameter 
describes the transitions in 1-type superconductors \cite{Evans95}
and in superfluidity $He$ - phases \cite{EPL1992}. Moreover, the
GL theory is exact in the symmetry aspects and will be used for
our following consideration of closed surface defects and its
evolution with connection to the shape of these surfaces.

\textbf{2. Topology of vesicular closed aggregates.} By studying
the membrane organization properties, Park et. al. \cite{EPL1992}
have revealed that the free-energy of a closed membrane (spherical
vesicle) is similar to the free-energy of a superconducting film
\cite{EPL1992}, \cite{Evans95}, and also to the free energy of a
transversal quantum magnetic flux with a superconducting vortex
flowing around (for the complex order parameter $\psi$). In this
analogy the magnetic flow corresponds to the director of some
smectic phase. The director defines the shape and stability
conditions of the vesicle \cite{EPL1992}. The GL free-energy
Hamiltonian with the complex order parameter $\psi$ is
\begin{eqnarray}\label{GL}
\textit{H}_{GL}=\int d^3x[r|\psi|^2+ C|(\nabla-ie^*\mathbf{A})\psi|^2+ \\
\frac{1}{2}u|\psi|^4
+\frac{1}{8\pi}(\nabla\times\mathbf{A}-\mathbf{H)^2] }.\nonumber
\end{eqnarray}
Here reduced charge $e^*=2e/\hbar c$, $\mathbf{H}$ is the magnetic
field and $\mathbf{A}$ is the vector potential. The order
parameter of the smectic phase $Sm-C$ depends on the number of
symmetry axis $n$ as $\psi=\langle exp[in\theta]\rangle$, being
$\theta$ an angle defining the orientation of the molecule tails,
$r$ and $C$ are expansion constants. Park et. al. applied the
Hamiltonian (\ref{GL}) to study closed equilibrium shapes with
spherical $n=0$ and ellipsoidal $n=1$ geometry, and also cases
with $n=$ 2, 3, 4, 6 symmetries. The Hamiltonian (\ref{GL})
describes second order phase transition between normal and
superconducting metals with Abrikosov vortex lattice phase with
finite density of vortexes $N_{\nu}$, $\int d^3x (\nabla \times
\mathbf{A})$ = $L N_{\nu} \phi_0$, where $L$ denotes the length of
the sample along the field direction $\mathbf{H}$ and
$\phi_0=hc/2e$. The transition between the symmetries of $n$-order
on closed surfaces is analogous to the transition in Abrikosov
phase with fixed number of vortexes.
\begin{eqnarray}\label{ord}
\psi=\psi_0\prod_{i=1}^{2n}(\sin\frac{\theta}{2}
\cos\frac{\theta_{1}}{2}\exp[i(\phi-\phi_i)/2] \nonumber \\
-\cos\frac{\theta}{2}\sin\frac{\theta_1}{2}\exp[-i(\phi-\phi_i)/2])
\equiv \psi_0 P(\Omega), \nonumber
\end{eqnarray}
where $\Omega_i=(\theta_i,\phi_i)$. The order parameter $\psi$ and
the shape parameter of the vesicle $\rho$ below the phase
transition temperature is calculated by minimizing $\textit{H}$
over $\phi_0$ and the zeros positions $\{\Omega\}$ of the function
$P(\Omega)$. For instance, in the case $n=3$ the polynomial
$P(\Omega)$ contains 81 terms. Without writing the expressions
\cite{EPL1992} corresponding to the effective density of free
energy as an expansion in Legendre polynomials, we recall that the
shape functions corresponding to a figure of $n$-atic order are of
the form $\rho(\Omega)=\psi_0^2\tilde{\rho}^{(n)}(\Omega)$, that
is, proportional to $\psi_0^2\sim r-r_c$.

This generate set has exactly $2n$ zeros at arbitrary positions on
the sphere and is equivalent in form to fractional wave functions
for anyons \cite{Haldane1}.

We can consider the quantum Hall wave functions, which describe
particles with fractional charge \cite{Haldane1}, \cite{Haldane2}.
The many-particle wave function is
\begin{equation}
\Psi_N^{(m)}=\prod_{i<j}(u_iv_j-u_jv_i)^m, 
\end{equation}
where $m$ is relative angular momentum of a pair of particles,
($(2l-m)$ is total angular momentum), $(u,v)=(\cos\frac{1}{2}\theta
e^{i\phi/2}, \sin\frac{1}{2}\theta e^{-i\phi/2})$ are spinor
variables describing the particle coordinates. The numerical
solution for three-particle wave function $\Psi_N^{(m)}$ in
spherical geometry was found in \cite{Haldane1}.

Taking into account the fluctuations effects, we will receive the
qualitatively changed results about the spherical functions
\cite{EPL1992}. The relation between the vesicle shape and the
evolution of surface defects could play a significant role.

\textbf{2.1. One example: Helfrich hat model with fractional
number of molecules.} In the context of the membrane organization
problems, Helfrich \cite{Helfr2001} considered fluctuations
effects on membrane surfaces. In his construction (so called "hat
model") he introduced the entropy term, dependent on the local
curvature, which is induced by the thermal individual molecules
motion, corresponding to de Broglie wavelengths. The entropy
\cite{Helfr2001} is
\begin{equation}\label{entrop}
S_0=-\frac{1}{2}k_B(\ln N+\ln (2\pi))
\end{equation}
and corresponds to an effective number of hats given by
\begin{equation}\label{Helf}
N_j=\frac{J_j^2(\textbf{r})dA}{J_j^2(\textbf{r})|_{max}A},
\end{equation}
however in entropy calculations this term plays the role of the
particle number and it follows from (\ref{Helf}), that is not an
integer number in general. In the expression (\ref{Helf})
$\textbf{r}=(x,y)$ are vector coordinates, $A$ is area under the
hat, or the area occupied by a molecule, $J_j$ is the mean
curvature.

This phenomenological entropy presentation serves to clarification
the physical reason of surface fluctuations, leading to the shape
transformation of whole vesicle.

\textbf{2.2. Vesicle symmetry and surface defects.} Among the
known types of topological defects (hedgehogs, boojums etc.) in
case of the closed membrane we have observed only surface ones. If
from the experimental data it is possible to restore distribution
of the director field $\mathbf {n}$, then it is possible to define
the characteristics of defects \cite{Volovik}, which in our case
are convenient for classifying as linear surface features
(boundaries of islands). They can be connected with the hedgehogs
who have left on a surface. In any case of the defects
displacement, the conservation law of a surface charge should be
complied out \cite{Volovik}.

Taking into account the Gauss-Bonnet formula, for the islands
boundaries we know the expression \cite{MM}:
\begin{equation}
\frac{1}{2\pi}d\theta+ \int \mathbf{w} (\mathbf{l} \times
\frac{d\mathbf{w} }{ds} )ds = m - \frac{1}{2\pi}
\int_\textit{u}KdS = m -\chi(\partial\textit{u}),
\end{equation}
where $\partial \textit{u}$ is the domain boundary, $\theta$ is
angle at going round the domain boundary,
 $\mathbf{l}$
is vortex vector,
$\mathbf{w}(\mathbf{l}\times\frac{d\mathbf{w}}{ds})=k_g$ is
geodesic curvature and $m$ is algebraic sum of dotty vortexes
circulation quanta in a domain $\textit{u}$. The surface topology
depends on the Euler characteristics $\chi$ and influence upon
total vortexes number \cite{MM}.

At the vesicle surface, the topological vortexes (with genus 0)
are directed both clockwise and counter-clockwise. Their number is
increasing in accordance to the vesicle sizes growth. The linear
vortexes are the boundaries of the surface islands. The number of
islands are always more, than one, when the simplex corresponds to
topologically regular partition \cite{MM}. In this case the next
equality take place: $E_+-E_-=\chi(M^2)-2N$, here $N$ is a number
of surface islands, $E_i$ - the number of positive and negative
directional simplex of partition $M^2$, as shown, it is whole here
and should correspond to the condensate with whole statistics.

\begin{figure}\label{figgg1}
\includegraphics*[width=80mm]{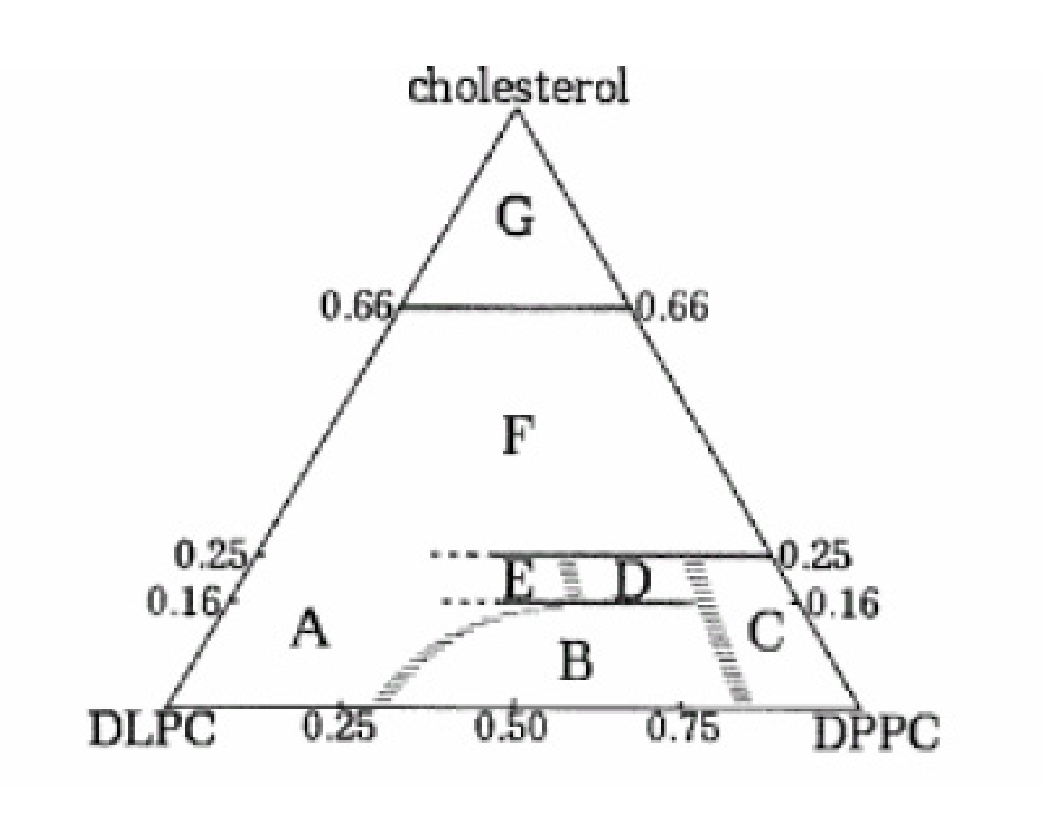}
\caption{\small Ternary phase diagram for $DPPC/DLPC/$cholesterol
at $297 K$ \cite{Feig}. Where A, $DLPC$-rich fluid lamellar phase;
B, coexisting fluid lamellar phase and $DPPC$-rich ordered phase;
C, $DPPC$-rich ordered phase; D, a single phase that changes
continuously from rigid ordered phase at the C/D boundary to a
fluid-ordered phase at the D/E boundary; E, a fluid-ordered phase;
F, a fluid-ordered phase different from E; G, coexisting
crystalline cholesterol monohydrate and a cholesterol-saturated
lamellar phase.}
\end{figure}

\begin{figure}\label{figgg2}
\includegraphics*[width=70mm]{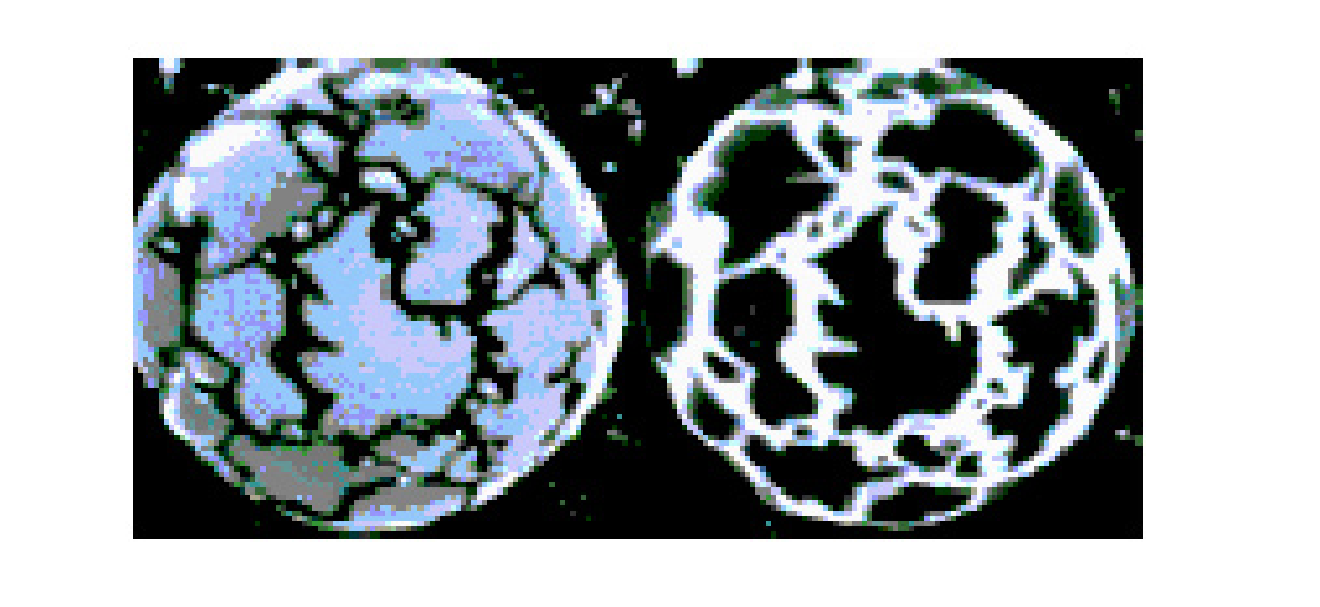}
\caption{\small The nanoscopic surface vesicle domains
\cite{Feig}.}
\end{figure}

At the spherical bilayer membrane, it is possible to observe two
types of regions: three-particles domains (consisting the
disordered mixture $DPPC/ DLPC$/cholesterol, these components are
labelling three different kinds of molecules) and one-particle
islands (with only lipids). Their common topological charge should
be the same at any lyotropic transformations. Looking onto the
evolution of surface domains \cite{Feig}, it is possible to
surmise, that the Euler characteristic become closed to two, but
nonequal it.

As may seem at first sight, at finite membrane thickness,
displacement and deformation of any defects are forbidden at all.
However, if the observed vesicle transformation, as expected
\cite{Feig}, is the continuous phase transition, then the
molecules of lipids come out from the volume always and they
supplement into the composition of the membrane or leave it.

The director field has a normal surface term, causing to outlet of
the hedgehogs to the surface. Topologically regular partition no
longer corresponds to the physical meaning of surface domains, the
Euler characteristic should be fractional.

The Laughlin wave functions of FQHE have not translational
invariance \cite{Haldane1}, but they are describing a circular
droplet of fluid, which must be confined in an external potential.

In \cite{Haldane2} for spherical geometry, the states with the
different defects combination, which correspond to Bose, Fermi and
fractional statistics, are described. But in thermodynamical limit
on a sphere, the particles of whole statistics stay only
\cite{Ei}. (Although there are manifolds (a torus \cite{Ei}),
where the particles of fractional statistics described by
multicomponent wave functions, are existing.)

Our membrane consist of the molecules of two amphiphilic
monolayers, whose tails are turned towards each other and heads
are directed toward the disordered phase inside and outside of the
bilayer sphere.

In our case of manifolds on two concentric spheres ($M_1 = S_1^2$
and $M_2=S_2^2$), both are connected by the value of the closed
membrane thickness, which is defined due to the lipid
conformation. The order parameters $\psi_1=\langle
\exp[in\theta_1]\rangle$ and $\psi_2=\langle
\exp[in\theta_2]\rangle$ are different, just as spinor variables
$(u_1,v_1)=(\cos\frac{1}{2}\theta_1 e^{i\phi_1/2},
\sin\frac{1}{2}\theta_1 e^{-i\phi_1/2})$ and for the second sphere
index, respectively. This fact of the connected state of the
closed spherical monolayers is an unproved hypothesis, but which
is permitted to explain the fractal axes and the asymmetric
vesicle shape.

\textbf{3. Concluding remark.} Thus, in light of FQHE description,
the each molecule of the same kind of lyotropic mixture may be
formally equivalent to anyon. Using a numerical simulation with
the statistical Monte Carlo algorithms, it will be possible to
define the type of observed phase transitions in actual and to
calculate the structural domain sizes suitable to the axes
geometry evolution.

The author have pleasure to thank M. I. Monastyrsky and O. P.
Santillan for the stimulant discussions. And the author grateful
to R. M. L. Evans for his reference to his paper. This research
was supported by Leading Scientific School of Russian Federation
(project 1907.2003.2.).

\end{document}